\documentclass[twocolumn,samsmath,amssymb,floatfix,superscriptaddress,nofootinbib]{revtex4}
\usepackage{graphicx}
\begin{document} 
\title{Liquid stability in a model for ortho-terphenyl}
\author{E.~La Nave}
\author{S.~Mossa}
\author{F.~Sciortino}
\author{P.~Tartaglia}
\affiliation{
Dipartimento di Fisica, INFM UdR and Center for Statistical 
Mechanics and Complexity,  Universit\`{a} di Roma ``La Sapienza'', 
P.le A. Moro 5, I-00185 Rome, Italy} 
\begin{abstract} 
We report an extensive study of the phase diagram of a 
simple model for ortho-terphenyl, focusing on the limits 
of stability of the liquid state. Reported data extend   
previous studies of  the same model to both lower and 
higher densities and to higher temperatures.  
We estimate the location of the homogeneous liquid-gas 
nucleation line and of the spinodal locus. Within the  
potential energy landscape formalism, we calculate the 
distributions of depth, number, and shape of the potential  
energy minima and show that the  statistical properties of 
the landscape are  consistent with a Gaussian distribution
of minima over a wide range of volumes. We report the volume dependence 
of the parameters entering in the Gaussian distribution 
(amplitude, average energy, variance). 
We finally evaluate the locus where the configurational entropy vanishes, 
the so-called Kauzmann line, and discuss the relative location of 
the spinodal and Kauzmann loci. 
\end{abstract} 
\maketitle 
\section{Introduction} 
\label{intro:sect}
Recent years have seen a strong development of numerical and theoretical 
studies of simple liquid models, attempting  to  develop a thermodynamic 
description based on formalisms which could be extended to deal also with 
out-of-equilibrium (glassy) 
states~\cite{speedy,kurchan,mezpar,angmk,teo,franz,leuzzi,mossa,soft}. 
Hard spheres, soft spheres, Lennard Jones mixtures and simple molecular 
liquids~\cite{speedybook,speedymolphys,voivod01,scott,starr01,sastry01,roberts,otplungo}
have been extensively studied. Within the potential energy landscape 
(PEL)~\cite{stillingerpes} thermodynamic approach, 
detailed comparisons between numerical data and theoretical predictions 
have been performed.  
Estimates of the number $\Omega$ of local minima --basins-- as a function 
of the basin depth and of their shape have  been recently evaluated 
for a few   
models~\cite{speedymolphys,skt99,scala00,starr01,sastry01,heuer00,voivod01,speedyjpc,otplungo}
and from the analysis of experimental data~\cite{stillinger98,richert98,speedyjpc2}. 
The PEL approach, which is particularly well suited for describing supercooled 
liquids, provides a controlled way to extrapolate the thermodynamic properties 
of the liquid state below the lowest temperature at which equilibrated 
data can be collected. For example, estimates of the locus where the 
configurational entropy vanishes, the so-called Kauzmann locus, can be
given~\cite{notakauz}; the Kauzmann locus provides a theoretical limit to the 
metastable liquid state at low  temperatures. 
Another limit to the liquid state is met on superheating and stretching 
the liquid, when the nucleation of the gas phase takes place.  
In this case a convenient way to define the limit of stability of 
the liquid state against gas nucleation is provided by the spinodal line, 
i.e., the locus of point where the compressibility diverges~\cite{pablobook}. 
The Kauzmann line and the spinodal line define the region of phase space 
where the liquid can exist in stable or metastable thermodynamic 
equilibrium.   
 
Recent theoretical and numerical work has focused on the thermodynamic relation 
between these two curves~\cite{kauzsdt,sriprl,soft,speedypisa,speedypreprint}.   
A recent thermodynamic analysis~\cite{speedypisa,speedypreprint} suggests that 
the spinodal and the Kauzmann loci meet --in the $(P,T)$ plane-- with the same 
slope at a point corresponding to the maximum tension that the supercooled 
liquid can sustain.   
Simple models which can be solved analytically, as hard~\cite{speedybook}  
or soft~\cite{speedypisa,scott} spheres complemented by a mean field attractive 
potential, support such prediction. A numerical study of a Lennard Jones mixture 
is also consistent~\cite{sriprl}.   

In this paper we consider the Lewis and Wahnstr\"{o}m rigid model for
the fragile glass former ortho-terphenyl (OTP)~\cite{lewis}, 
whose dynamic~\cite{lewis,rinaldi,chong} and thermodynamic 
features~\cite{otplungo,lanave} have been studied in detail. 
Our aim is to calculate the spinodal and the Kauzmann 
lines to estimate the region of stability of the liquid, and study
the relation between these two loci. 

We improve the data base of phase state points of previous 
studies~\cite{otplungo}, extending to both lower and higher densities and to
higher temperatures. Performing analysis of the pressure-volume 
relation along several isotherms, we estimate the homogeneous nucleation line 
and the spinodal curve; we also report upgraded estimates 
of the statistical properties of the landscape sampled by the liquid. 
We confirm that at all densities a Gaussian landscape properly models the 
thermodynamics of the system in the supercooled state. Such agreement gives us 
confidence in the evaluation of the locus along which the configurational 
entropy vanishes. We finally discuss the limits and possibilities of an analysis 
based on the inherent structures thermodynamic formalism. 
The description of our results is preceded by a short review of the potential 
energy landscape approach to the thermodynamics of supercooled liquids. 
\section{Background: the free energy in the inherent structures thermodynamic 
formalism} 
\label{background1:sect}
An expression for the liquid free energy in the range of temperatures 
where the liquid is supercooled can be given within the PEL formalism. 
In supercooled states, i.e., when the correlation functions 
show the two steps behavior typical of the cage
effect~\cite{kob-review,mct1}, the system's properties are controlled 
by the statistical properties of the PEL~\cite{sds}.
In the PEL formalism, the potential energy hyper-surface 
--fixed at constant volume--
is partitioned into basins; each basin is defined as the set of 
points such that a steepest descent path originating from them ends in 
the same local minimum. The configuration corresponding to the minimum is called 
inherent structure (IS), of energy $e_{IS}$ and pressure $P_{IS}$.  
The partition function can be expressed as the sum over all the basins,
weighted by the appropriate Boltzmann factor, i.e., as a sum over the single 
basin's partition functions.  
As a result, the Helmholtz liquid free energy $F(T,V)$, at temperature $T$ and
volume $V$, can be written as~\cite{stillingerpes}:  
\begin{equation} 
F(T,V)=  E_{IS}(T,V) - T S_{conf} (T,V)+ f_{vib} (T,V). 
\label{eq:freeenergy} 
\end{equation} 
Here, $E_{IS}$ is the average energy of the IS explored at the given $(T,V)$,  
$f_{vib}$ is the vibrational free energy, i.e., the average free  energy of 
the system when constrained in a basin of depth $e_{IS}$, 
and $S_{conf}(T,V)$ is the configurational entropy, that counts the number  
of explored basins. $S_{conf}(T,V)$ is a quantity of crucial interest,
both for comparing numerical results with the recent theoretical 
calculations~\cite{mezpar,coluzzi}, and to examine some of the proposed relations 
between dynamics and thermodynamics~\cite{adam65,schultz,wolynes}. 
Therefore, in order to evaluate the free energy one needs to estimate 
the three terms of Eq.~(\ref{eq:freeenergy}).   
$E_{IS}(T,V)$ is calculated by means of a steepest descent potential energy 
local minimization of equilibrium configurations (see Ref.~\cite{otppisa} 
for details). In fragile liquids the $T$ dependence of $E_{IS}(T,V)$ 
follows a $1/T$ law~\cite{heuer,sastry01,starr01,otplungo}.   

The basin free energy $f_{vib}(T,V)$ takes into account 
both the basin's shape --curvature-- and the system kinetic energy. 
From the formal point of view, this term is the integral of the Boltzmann  
factor constrained in a basin, averaged over all the basins with same depth 
$e_{IS}(T,V)$. 
Numerically, $f_{vib}$ is evaluated  as sum of two contributions: 
{\em i)} a harmonic contribution, which depends on the curvature of the 
accessed basins corresponding to the minimum at the given $e_{IS}$;  
{\em ii)} an anharmonic contribution, usually approximated as a function  
of $T$ only. 
In the case of a rigid molecule model, the harmonic contribution is the free energy 
associated with a system of $(6N-3)$ independent oscillators of frequency $\omega_k$, 
where the  $\omega_k$ are the square root of the eigenvalues of the system 
Hessian matrix --the matrix of the second derivatives of the potential energy-- 
evaluated at the basin minimum (see Ref.~\cite{otppisa} for details). 
This contribution can be written as~\cite{notaerrore}:
\begin{equation}
f_{harm}(E_{IS},T,V)=
\ln \langle \exp\left\{\sum_{i=1}^{6N-3}\ln(\beta\hbar\omega_i(e_{IS}))\right\} 
\rangle',
\label{eq:fvib}
\end{equation}
where the symbol $\langle  \rangle'$ denotes the average over all the 
basins with the same energy $e_{IS}$.  
The $E_{IS}$ dependence of $f_{vib}$, in the harmonic approximation, can be  
parameterized using the expression: 
\begin{equation} 
\ln \langle \exp\left\{\sum_{i=1}^{6N-3}\ln(\beta\hbar\omega_i(E_{IS}))\right\} 
\rangle'=a(V)+b(V) E_{IS} + c(V) E_{IS}^2. 
\label{eq:shape} 
\end{equation} 
The evaluation of $f_{anh}(T,V)$ is described in Ref.~\cite{otppisa}.

Finally, $S_{conf}$ can be calculated as the difference of the (total) entropic 
part of Eq.~(\ref{eq:freeenergy}) and the vibrational contribution to the entropy: 
\begin{equation} 
S_{conf} (T,V) = S (T,V) - S_{\rm harm} (T,V) - S_{\rm anh} (T,V). 
\label{eq:sconf} 
\end{equation} 
\section{Background: the Gaussian landscape} 
\label{background2:sect}
In order to evaluate analitically the free energy in the PEL formalism, 
it is necessary to provide a model for the probability distribution 
of $e_{IS}$ , i.e., the number $\Omega(e_{IS}) de_{IS}$ of basins 
whose depth lies between $e_{IS}$ and $e_{IS} + de_{IS}$.   
Among several possibilities, the Gaussian 
distribution~\cite{rem,heuer00,sastrynature} seems to provide a satisfactory 
description of the numerical simulations of 
Refs.~\cite{sastrynature,otplungo,starr01}. 
The ``Gaussian landscape'' is defined by:
\begin{equation} 
\Omega(e_{IS})de_{IS}=e^{\alpha N} 
\frac{e^{-(e_{IS}-E_o)^2/2\sigma^2}}{(2 \sigma^2)^{1/2}}de_{IS}. 
\label{eq:gdos} 
\end{equation} 
Here, the amplitude $e^{\alpha N}$ accounts for the total number of 
basins, $E_o$ plays the role of energy scale and $\sigma^2$,
measures the width of the distribution. One can grasp the origin of such 
distribution invoking the central limit theorem. 
Indeed, in the absence of a diverging correlation length, in the thermodynamic limit,  
each IS can be decomposed in a sum of independent subsystems~\cite{notadistribuzione},  
each of them characterized by its own value of $e_{IS}$. 
The IS energy of the entire system, in this case, will be distributed
according to Eq.~(\ref{eq:gdos}).

The  assumptions of a Gaussian Landscape (Eq.~(\ref{eq:gdos})) 
and of a  quadratic dependence of the basin free energy on  $E_{IS}$  
(Eq.~(\ref{eq:shape})) fully specify the statistical properties of
the model. Thus, it is possible to evaluate the $T$ dependence of  
$E_{IS}$ and $S_{conf}$. The corresponding expressions are~\cite{otppisa}: 
\begin{equation} 
E_{IS}(T,V) =\frac{(E_o(V)-(b(V)+\beta)\sigma^2(V))} 
{1+2 c(V) \sigma^2(V)}=A+\frac{B}{T}, 
\label{eq:eis}  
\end{equation} 
where, for convenience, we have defined $A=\frac{E_o-b \sigma^2}
{1+2 c \sigma^2}$ and $B=\frac{-\sigma^2}{k_B (1+2 c \sigma^2)}$; and
\begin{equation} 
S_{conf}(T,V)/k_B=\alpha(V) N -\frac{(E_{IS}(T,V) -E_o(V))^2}{2
\sigma^2(V)}.
\label{eq:sconf2} 
\end{equation} 
Note that $E_{IS}$ is linear in $1/T$. The predicted $1/T$ dependence  
of $E_{IS}$ and the parabolic dependence of $S_{conf}$ has been confirmed in 
several models for fragile liquids~\cite{sastrynature,otplungo,starr01}. 

From the relations above and fits of the numerical data, one obtains:
{\em i)} the vibrational coefficients $a$, $b$ and $c$ from 
Eq.~(\ref{eq:shape}); {\em ii)} the distribution parameters, $E_o$ 
and $\sigma^2(V)$ from Eq.~(\ref{eq:eis});  
{\em iii)} the amplitude $e^{\alpha N}$ from Eq.~(\ref{eq:sconf2}).
A study of the volume dependence of the parameters $\alpha(V)$, $E_o(V)$, 
and $\sigma^2(V)$ , associated with the $V$-dependence 
of the shape indicators (Eq.~(\ref{eq:shape})), provides a full 
characterization of the volume dependence of the landscape properties of a model, 
and offers the possibility of developing an equation of state based on
the volume dependence of the statistical properties of the landscape~\cite{lanave}. 

Finally, within the Gaussian landscape model, it is possible to exactly  
evaluate the Kauzmann curve $T_K(T,V)$ , the limit for the existence of the  
liquid. This curve is  the  locus of  points where $S_{conf}$ 
vanishes, i.e., from Eq.~(\ref{eq:sconf2}),
\begin{equation} 
\alpha(V) N 
-\frac{(E_{IS}(T,V) -E_o(V))^2}{2\sigma^2(V)} =0.
\label{eq:sconf3} 
\end{equation}
The following expression for $T_K(T,V)$ results: 
\begin{equation} 
T_K(T,V)=\frac{B}{(E_o-A) \pm \sqrt{2 \sigma^2 \alpha N}},
\label{eqtk} 
\end{equation} 
where the sign to be chosen is the one corresponding to the largest value 
solution of the equation. 
\begin{figure}[t]
\centering
\includegraphics[width=0.48\textwidth]{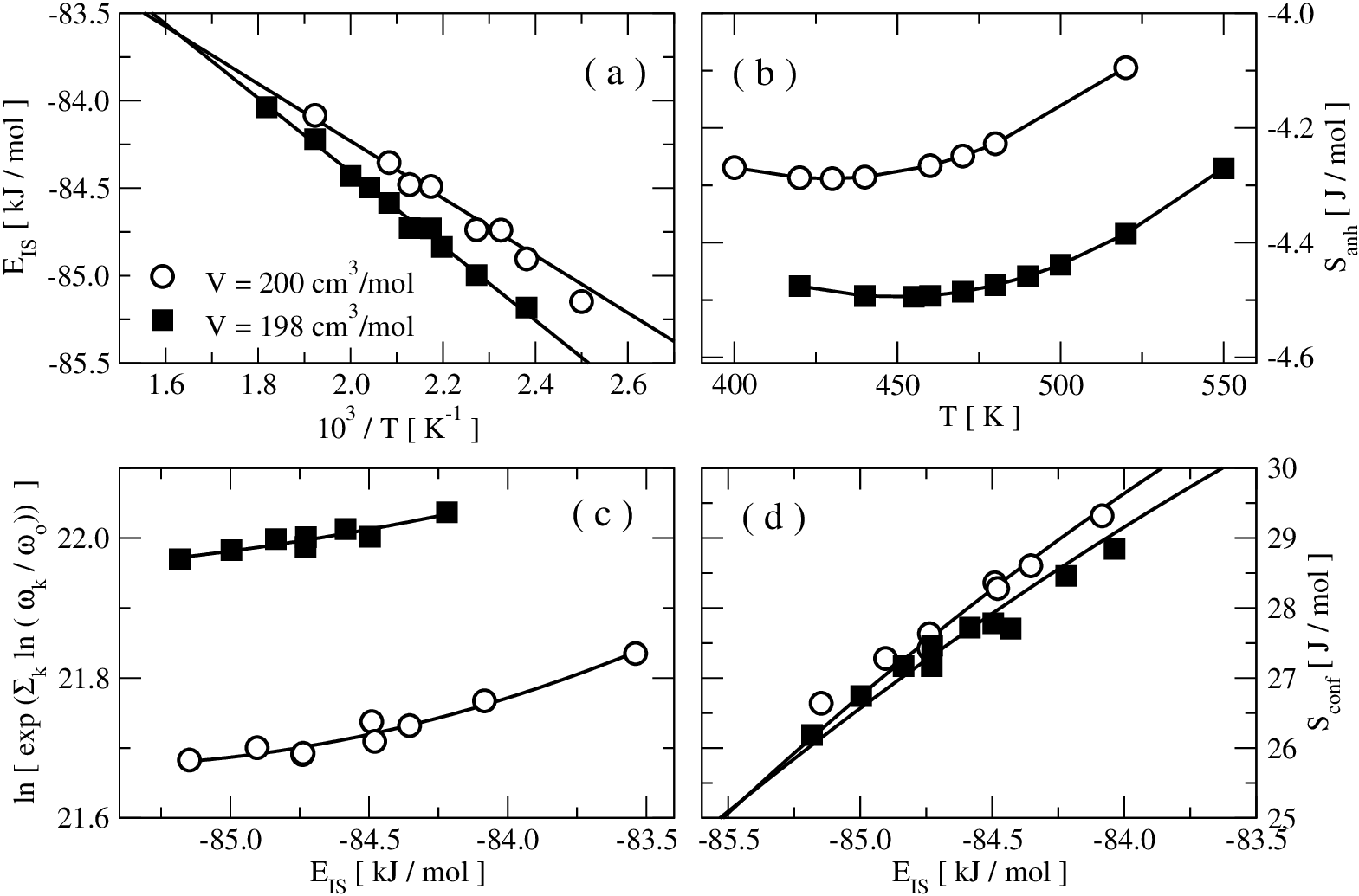}
\caption{Results of the landscape analysis for the  
densities $198$ and $200$ cm$^3$/mol (squares and 
circles symbols respectively): {\em a)} average inherent structure energy
(symbols) together with the $1/T$ dependence of 
Eq.~(\protect\ref{eq:eis}) (solid line), 
{\em b)} anharmonic contribution to the entropy,
{\em c)} harmonic free energy $\ln\langle \exp\left\{\sum_{i=1}^{6N-3}
\ln(\beta\hbar\omega_i(E_{IS}))\right\} \rangle$, 
{\em d)} configurational entropy $S_{conf}$ together with the fits
according to Eq.~(\protect\ref{eq:sconf2}).}
\label{fig:all}
\end{figure}
\section{Model and simulations}
\label{model}
We studied a system of $343$ Lewis and Wahnstr\"{o}m (LW)~\cite{lewis} 
ortho-terphenyl model molecules, by means of molecular dynamics
simulations in the $(N,V,T)$ ensemble.
The LW model is a rigid, three-site model with intermolecular 
site-site interactions described by  Lennard
Jones potential. The potential parameters are chosen to  reproduce  
OTP  properties such as its structure and diffusion coefficient~\cite{lewis}. 
The integration time step for the simulation was $0.01$ ps. 
With this model it is possible to reach very long simulation times;
such long molecular dynamics trajectories allow us to equilibrate the system 
at temperatures below the temperature where the diffusion constant reaches 
values of order $10^{-10}$ cm$^2$/s.
We simulated  $23$ different densities for several temperatures, 
for an overall simulation time of order $10$ $\mu$s.
\begin{figure}[t]
\centering
\includegraphics[width=0.4\textwidth]{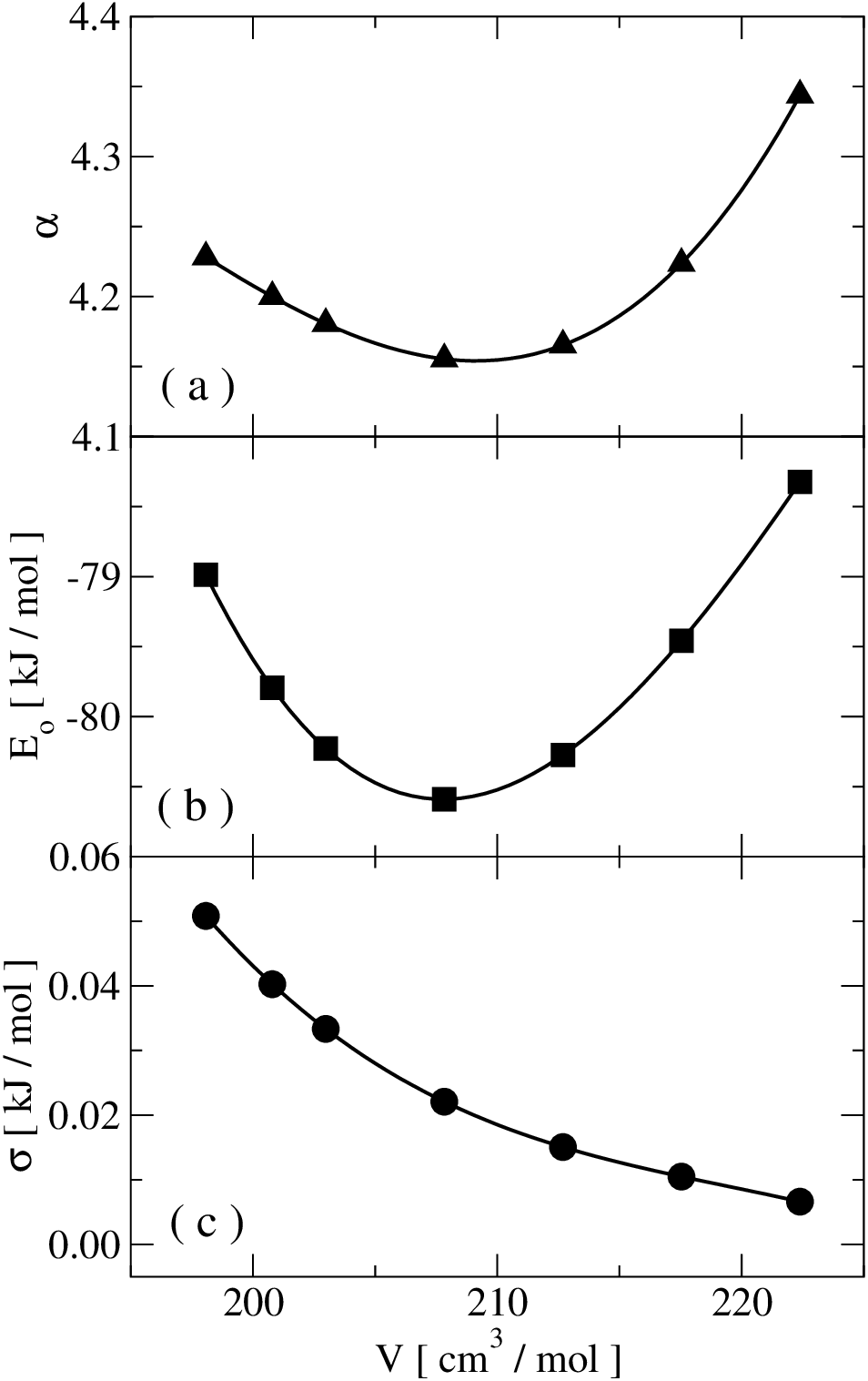}
\caption{Volume dependence of the Gaussian landscape parameters:  
{\em a)} $\alpha(V)$, {\em b)} $E_o(V)$, and {\em c)} $\sigma^2(V)$.}
\label{parmfis}
\end{figure}

To calculate the inherent structures sampled by the system in equilibrium 
we perform conjugate gradient energy minimizations to locate the closest local 
minima on the PEL, with a tolerance of $10^{-15}$ kJ/mol.
For each thermodynamical point we minimize at least $100$ configurations,  
and we diagonalize the Hessian matrix of at least $50$ configurations  
to determine the density of states. The Hessian is calculated choosing 
for each molecule  the center of mass and the angles associated with 
rotations around the three principal inertia axis as coordinates.

Further details on the numerical techniques used can be found 
in Ref.~\cite{otplungo}.
\begin{figure}[t]
\centering
\includegraphics[width=0.45\textwidth]{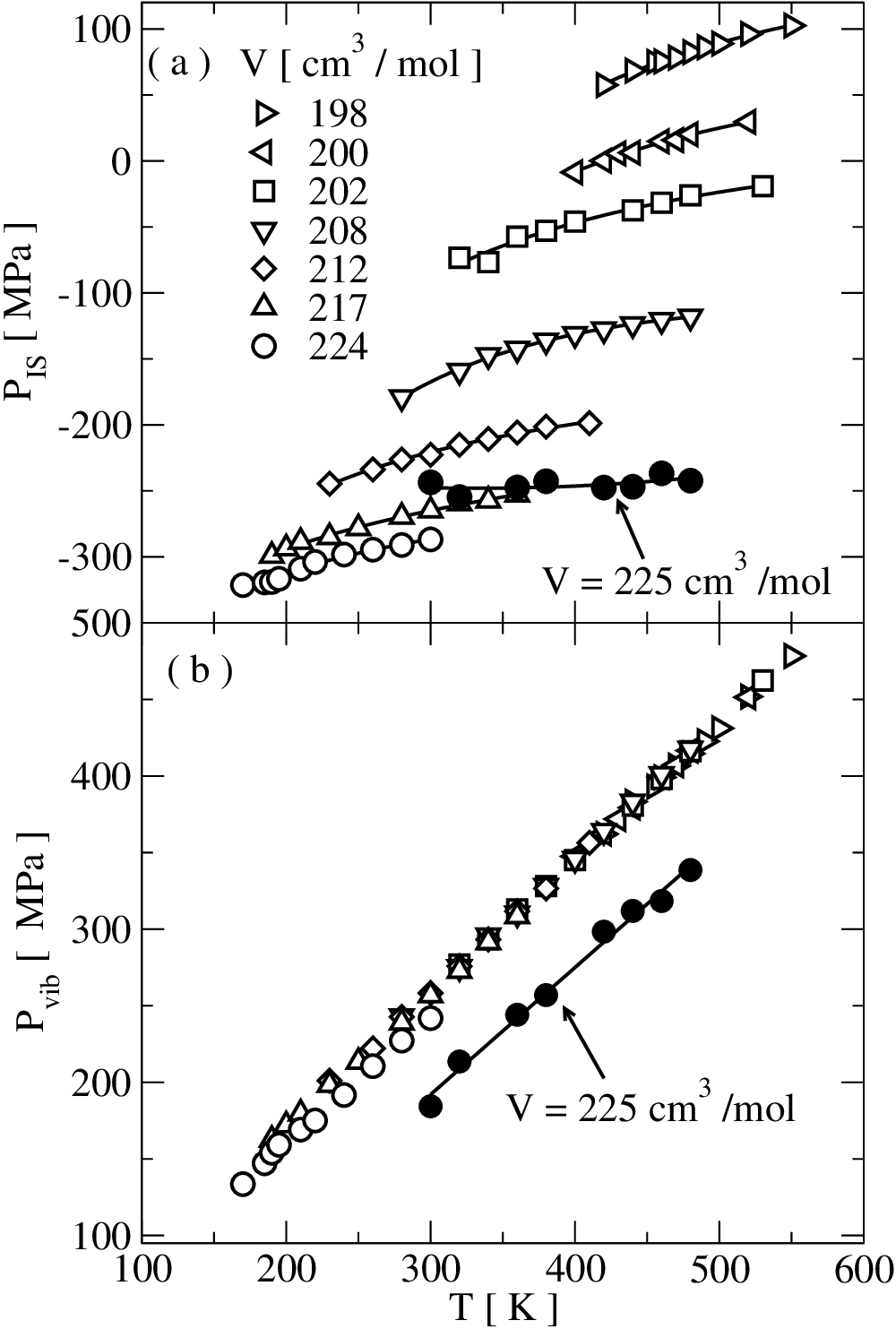}
\caption{{\em a)} Inherent structure pressure, 
$P_{IS}$. The IS pressure gets more negative on 
decreasing density, until a sudden jump upward takes place at 
around $225$ cm$^3$/mol. 
The jump signals that, during the minimization procedure 
(at constant volume), the maximum tensile strength has been overcome 
and a cavitation phenomenon has taken place. 
{\em b)} vibrational pressure, $P_{vib}$. $P_{vib}$ is almost density 
independent. Only at densities close to the Sastry density, a $V$ 
dependence is observed.}
\label{fig:pis}
\end{figure}
\section{Potential energy landscape properties} 
\label{landscape:sect}
An analysis of the statistical properties of the landscape 
for the LW ortho-terphenyl model has been recently performed
in Ref.~\cite{otplungo}. 
Here we expand such analysis to lower and higher densities,
with the aim of exploring the region of phase diagram where
the steep repulsive part of the potential is more relevant, 
and study the location of the liquid spinodal line and of the 
Sastry density~\cite{sripre,sriprl}. 
The larger density range considered allows us to estimate 
the volume dependence of the landscape parameters with great precision. 
\begin{figure}[t]
\centering
\includegraphics[width=0.48\textwidth]{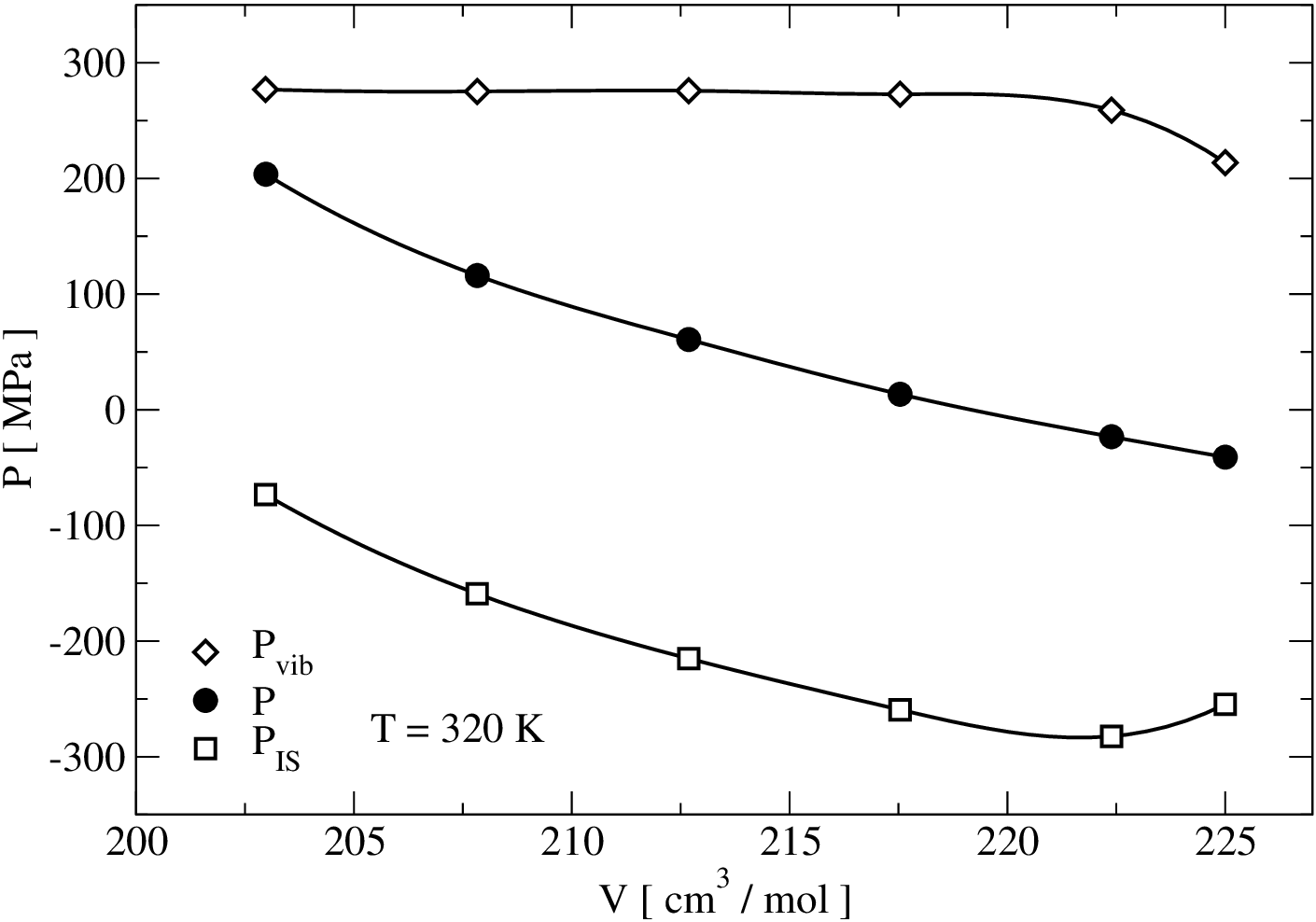}
\caption{Volume dependence of the total pressure $P$, 
inherent structures pressure $P_{IS}$, and vibrational pressure 
$P_{vib}$ for the isotherm $T=320$ K.
The total pressure $P$ is monotonically decreasing, as expected for a system
in equilibrium in the stable liquid phase. 
The IS pressure shows instead a minimum around $V=222$ cm$^3$/mol, 
suggesting that, during the minimization process, 
a cavitation process has taken place in the $V=225$ cm$^3$/mol sample. 
As a result, the significant drop observed in $P_{vib}$ is an artifact 
induced by cavitation.}
\label{fig:pvc}
\end{figure}

The four panels of Fig.~\ref{fig:all} show the results of the landscape 
analysis for two of the additional densities ($198$ and $200$  cm$^3$/mol), 
with the aim of confirming the possibility of describing the numerical data 
with the Gaussian landscape model discussed above. 
Similar data for other five densities can be found in 
Ref.~\cite{otplungo} and are not shown here. 
Fig.~\ref{fig:all}(a) shows the temperature dependence of the average 
inherent structure energy (symbols) which, in agreement with 
Eq.~(\ref{eq:eis}), can well be fitted by a $1/T$ law (solid line).  
Fig.~\ref{fig:all}(b) shows the anharmonic entropy,  
evaluated according to a fit of the anharmonic contribution to the energy 
with a polynomial of third degree  in $T$~\cite{otplungo}.  
Fig.~\ref{fig:all}(c) shows the quantity 
$\ln\langle\exp\left\{\sum_{i=1}^{6N-3}\ln(\beta\hbar\omega_i(E_{IS}))\right\} 
\rangle'$ as a function of the basin depth energy $E_{IS}$.  
As previously observed, an almost linear relation between  
$\ln\langle\exp\left\{\sum_{i=1}^{6N-3}\ln(\beta\hbar\omega_i(E_{IS}))\right\}
\rangle'$ and $E_{IS}$ is found. To account for a minor curvature, 
a fit with a second order polynomial (Eq.~\ref{eq:shape}) is reported. 
Finally, the $E_{IS}$ dependence of  
$S_{conf}$ is shown in Fig.~\ref{fig:all}(d).   
The parameters of the reported fit are constrained 
by the $T-$dependence  of the parameters estimated for 
$E_{IS}$~\cite{otppisa}. In agreement with Eq.s~(\ref{eq:eis}) 
and (\ref{eq:sconf2}), $S_{conf}$ is fitted by a second degree 
polynomial in $E_{IS}$.
 
As discussed in the previous section, from the $T$ dependence of $E_{IS}$ 
and $S_{conf}$, and from the $E_{IS}$ dependence of 
$\ln\langle \exp\left\{\sum_{i=1}^{6N-3}
\ln(\beta\hbar\omega_i(E_{IS}))\right\}\rangle'$, it is possible 
to evaluate the statistical properties of the landscape and their volume 
dependence, under the assumption of a Gaussian landscape.  
Fig.~\ref{parmfis} shows the $V$-dependence of the parameters  
$\alpha(V)$, $E_o(V)$, and $\sigma^2(V)$. The parameter $\alpha$ shows 
a weak $V$-dependence. From a theoretical point of view, we expect 
$\alpha$ to converge toward a constant value in the small 
volume limit, when the potential is essentially dominated by the repulsive 
soft sphere $r^{-12}$ part~\cite{scott}, and to increase on increasing $V$,  
to account for the larger configuration space volume. The $V$-dependence 
of $E_o(V)$ is similar to the one found in other models;
the presence of the minimum is, indeed, connected to the progressive 
sampling, on compression, of the attractive part of the intramolecular 
potential, followed by the progressive probing of the repulsive part 
of the potential. As expected for simple liquids~\cite{spceprl}, 
the $V$-dependence of  $\sigma^2(V)$ is  instead monotonic. 
\begin{figure}[t] 
\centering
\includegraphics[width=0.48\textwidth]{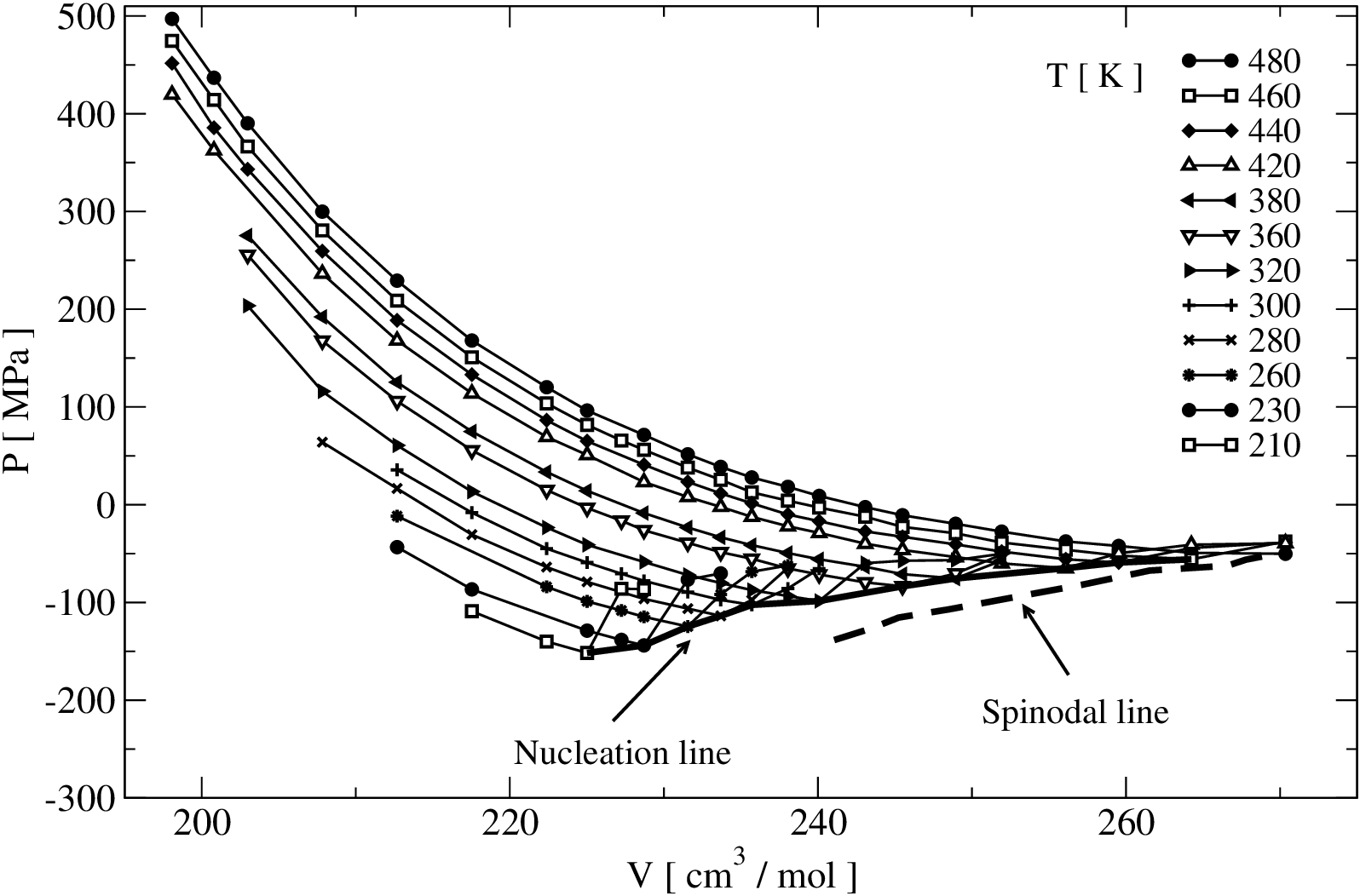} 
\caption{Total system pressure $P$ as a function of the volume $V$ along 
the indicated isotherms. For each isotherm, the highest volume shown is the 
volume where a jump in the inherent structure pressure is first observed, 
signaling bubble formation --cavitation-- in the system.
The nucleation line (solid line) and the spinodal line (dashed line)
are also shown.} 
\label{fig:pt} 
\end{figure} 
\section{Inherent Structures Pressure, $P_{IS}$, and vibrational 
contribution, $P_{vib}$} 
\label{sec:eos} 
Within the landscape approach, the pressure $P$ can be 
exactly split in two contributions, one associated to 
the pressure  $P_{IS}$ experienced in the local minima  
(which are usually under tensile or compression stress), 
and a second contribution, $P_{vib}$,  
commonly named vibrational, even if a configurational part 
is also included as discussed in length in Ref.~\cite{scott,soft}. 
Hence, in full generality,
\begin{equation} 
P(T,V)=P_{IS}(T,V)+P_{vib}(T,V). 
\label{eq:ptotal}
\end{equation} 
Here, $P_{IS}(T,V)$ can be evaluated from the value of the virial 
expression in the inherent structure, while $P_{vib}(T,V)$ can be 
evaluated as difference between $P$ and $P_{IS}$.  

Fig.~\ref{fig:pis}(a) shows the inherent structure pressure 
$P_{IS}$ and  Fig.\ref{fig:pis}(b) the vibrational pressure $P_{vib}$   
for several densities. The IS pressure gets more negative on 
decreasing density, until a sudden jump upward takes place at 
around $225$ cm$^3$/mol. 
The jump signals that, during the minimization procedure 
(at constant volume), the maximum tensile strength has been overcome 
and a cavitation phenomenon has taken place. 
Therefore, the density at which the IS loses mechanical stability, 
recently named Sastry density~\cite{debenlewis} ($V_{Sastry}$), is, 
for this model, close to $224$ cm$^3$/mol. 
We also note that $P_{vib}$ (Fig.~\ref{fig:pis}(b)) is almost density 
independent. 
Only at densities close to the Sastry density, a $V$ dependence 
is observed. For densities lower than the Sastry density, both $P_{IS}$ 
and $P_{vib}$ do not reflect any longer bulk properties 
(being the local minima configuration affected by the presence of large voids).
\begin{figure}[t] 
\centering
\includegraphics[width=0.48\textwidth]{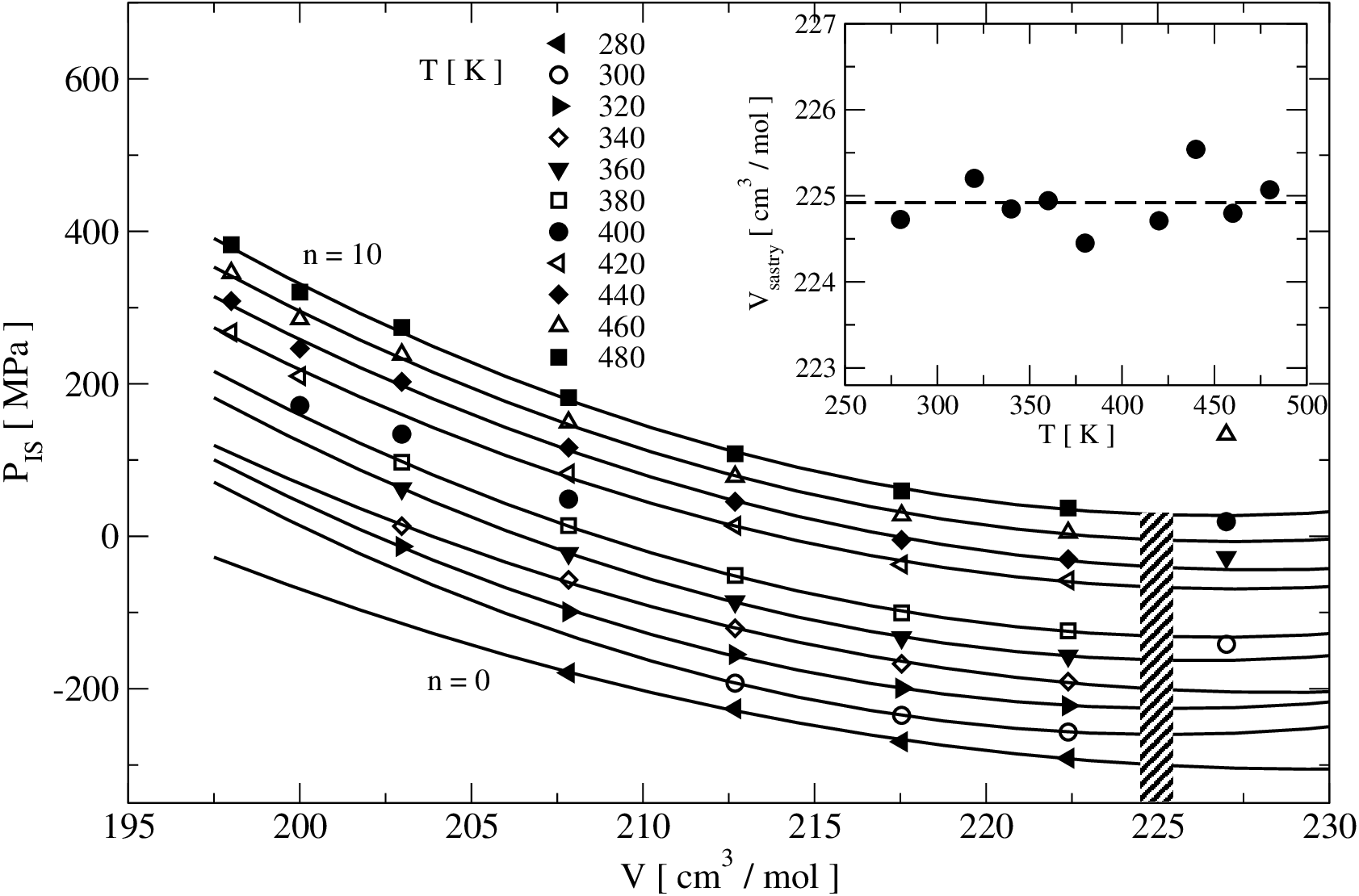} 
\caption{Inherent structures pressure, $P_{IS}$, as a function of $V$, 
for several isotherms. Each curve has been shifted by $n \times 30$ MPa,
for a more clear presentation.
Solid lines are the theoretical predictions according to the potential energy
landscape equation of state of Ref.~\protect\cite{lanave}.
The shadowed region marks the range of variability of $V_{Sastry}$, 
the limit of stability for the inherent structures. 
Inset: $V_{Sastry}$ as a function of temperature for all the studied 
volumes is shown. Dashed line is the average value $224.9$ cm$^3$/mol.} 
\label{fig:pis2} 
\end{figure} 

The previous observation stands out more clearly in Fig.~\ref{fig:pvc}
where the volume dependence of $P$, $P_{IS}$ and $P_{vib}$ is shown
for the isotherm $T=320$ K. The total pressure $P$ is
monotonically decreasing, confirming that the studied system is in the 
stable liquid phase up to $V=225$ cm$^3$/mol.  The IS pressure
shows instead a minimum around $V=222$ cm$^3$/mol, suggesting that, 
during the minimization process, a small cavity has been created in the  
$V=225$ cm$^3$/mol sample. 
As a result, the significant drop observed at $V=225$ cm$^3$/mol 
in $P_{vib}$ is an artifact induced by cavitation. It is also worth 
noting that a small decrease of $P_{vib}$ is observed already at $V=222$ 
cm$^3$/mol, suggesting that the cavitation phenomenon is preceded by a 
weak softening of the vibrational density of states on approaching 
the Sastry instability. 
\begin{figure}[t] 
\centering
\includegraphics[width=0.48\textwidth]{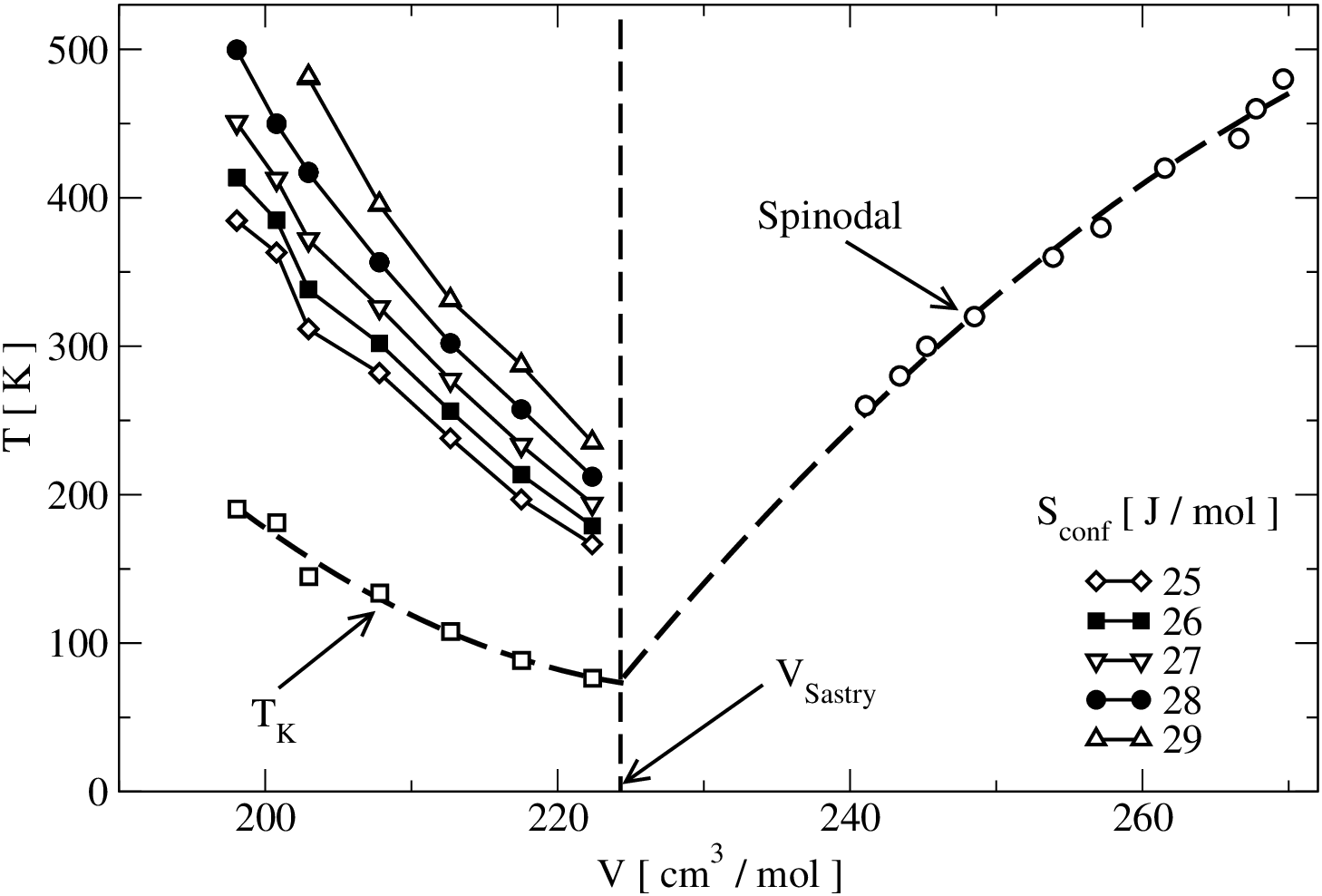}
\caption{Spinodal (open squares) and Kauzmann (open circles) lines,
together with their extrapolations (dashed lines) in the $(T,V)$ plane. 
The relation between these two curves is discussed in the text.
The vertical dashed line marks the Sastry volume, the limit of stability
for the inherent structures. Some iso-entropic lines are also shown.} 
\label{fig:locivt} 
\end{figure}
\section{Limits of stability of the liquid} 
\label{sec:stability}
We now focus on the limits of stability of the liquid state. 
Fig.~\ref{fig:pt} shows the volume dependence of the pressure 
for several of the studied isotherms. Cavitation marks the 
homogeneous nucleation limit for the system; 
it can be detected during the simulation by monitoring the time 
dependence of pressure and potential energy, which show a clear 
discontinuity when a gas bubble nucleates. 
After the cavitation, $P$ increases and the potential energy decreases.   
We define the locus of homogeneous nucleation as the
largest volume, at fixed $T$, at which we managed to 
simulate the dynamics of a homogeneous system.  
The fine grid of studied $V$ values allows us 
to identify such locus with a significant precision (solid line). 
Although the calculated  homogeneous nucleation line refers to a system 
composed of $343$ molecules, it provides an upper bound for larger 
systems.

From the equilibrium $P(T,V)$ data, i.e., in the range where no 
cavitation is observed, it is possible to estimate the volume  
at which $P(V)$ has a minimum, by fitting the data according to the 
equation $P \approx  (V-V_s)^2$. 
In the mean field approximation, $V_s$ corresponds to the spinodal 
volume, and the temperature dependence of $V_s$ defines the spinodal 
locus (dashed line).
\begin{figure}[t]
\centering
\includegraphics[width=0.48\textwidth]{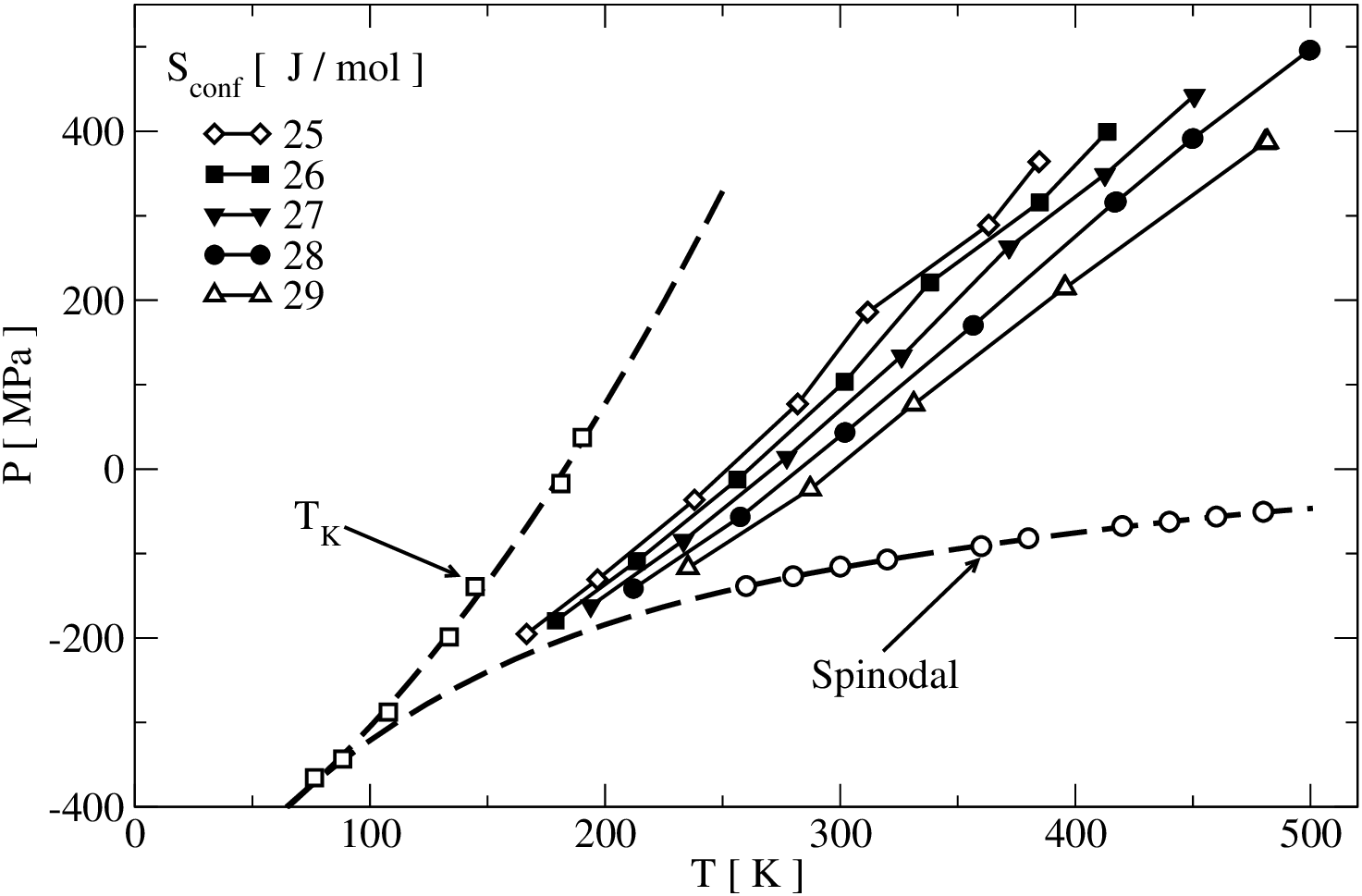} 
\caption{Spinodal (open squares) and Kauzmann (open circles) lines,
together with their extrapolations (dashed lines) in the $(P,T)$ plane. 
The vertical dashed line marks the Sastry volume. 
Some iso-entropic lines are also shown.} 
\label{fig:locitp}
\end{figure}

We now focus on the information on liquid stability encoded in 
the inherent structures pressure $P_{IS}$. 
Fig.~\ref{fig:pis2} shows, for several isotherms, $P_{IS}$ 
as a function of $V$. In analogy with the equilibrium data, a limit of 
stability for  the inherent structures can be calculated estimating the 
minimum of $P_{IS} \approx (V-V_{Sastry})^2$ (fits are solid lines). 
It has been speculated that $V_{Sastry}$ represent the upper  
limit for glass formation, and that it may be identified  with the 
$T \rightarrow 0$ limit of the liquid-gas spinodal locus~\cite{sripre}. 
Our calculations show that, as previously observed for other 
systems~\cite{sriprl}, the Sastry volume does not depend significantly 
on the temperature (see inset in Fig.~\ref{fig:pis2}).

We note that the Sastry volume is significantly smaller than the spinodal 
volume. The fact that $V_{Sastry} < V_{s}$  implies a stabilizing role 
of the vibrational component of the pressure. 
Indeed, already in  Fig.~\ref{fig:pvc} it was shown that the stability 
region for $P$ is larger than the one for $P_{IS}$. 
This fact suggests that, close to the spinodal line, the vibrational 
component becomes volume dependent to compensate for the loss of 
stability arising from the $P_{IS}$ contribution.  
Clearly, if $P_{vib}$ were $V$ independent, 
then $V_s$ and $V_{Sastry}$ should coincide. 
There is also an interesting observation to make concerning the 
application of the potential energy landscape approach to liquids at 
large volumes.  
Indeed, between $V_{Sastry}$ and $V_s$, the constant volume minimization 
procedure produces an inhomogeneous IS configuration~\cite{notasastry}.
When the inherent structure contains voids, the determination of the 
landscape parameters proposed in this work becomes meaningless,  
and the link between the inherent structure and the corresponding liquid 
state requires a more detailed modeling. 
 
Next, we focus on the location of the relevant stability loci of the liquid 
phase under  supercooling.  
Within the Gaussian landscape model,  the limit of stability of the supercooled 
liquid is defined by the line at which the configurational entropy 
vanishes, the so-called Kauzmann locus (Eq.~(\ref{eqtk})).  
As the spinodal curve is preempted by the homogeneous nucleation and, 
therefore, can not be approached in equilibrium, the Kauzmann locus can 
not be accessed due to the extremely slow structural relaxation times 
close to it. Still, these two loci provide a characterization of the domain 
of stability of the liquid state, and retain some meaning in a limiting 
mean-field sense. 
 
Figs.~\ref{fig:locivt} and \ref{fig:locitp} show the $S_{conf}=0$ locus 
and the spinodal locus in the $(T,V)$ and $(P,T)$ planes.   
The $(P,T)$ data have been calculated using the potential energy
landscape equation of state introduced in Ref.~\cite{lanave}. 
Fig.~\ref{fig:locivt} also shows curves at constant 
configurational entropy, in the region where no extrapolations are 
required. In analogy with the findings in Lennard Jones systems~\cite{sriprl}, 
the volume where the two loci appear to meet is close to $V_{Sastry}$, but 
at a temperature $T_I \simeq  73 K $ different from $T=0$. 
Hence, the spinodal line terminates at a finite temperature, by merging 
with the $S_{conf}=0$ line; this is at odds with what was suggested in 
Ref.~\cite{sripre}.  
For $T _I\simeq 73 K$, the glass will meet a mechanical instability on 
stretching; it is an interesting topic of 
research~\cite{angelljcp,angellprb,sriprl} to understand 
the relations between the volume at which such instability 
takes place and $V_{Sastry}$.	

A recent  thermodynamic analysis  addresses  the issue of the
relative location of the two loci and the way these two loci 
intersect~\cite{speedypisa,speedypreprint}.   
It has been suggested that the two lines must meet in $(P,T)$  plane 
with the same slope. 
Making use of the equation of state of Ref.~\cite{lanave}, 
the $(T,V)$ data of Fig.~\ref{fig:locivt} can be represented in the 
$(P,T)$ plane. The  phase diagram in $(P,T)$ is shown in 
Fig.~\ref{fig:locitp}, where also $T_K(P)$, the spinodal line and 
some iso-entropy curves are shown.
An extrapolation of the low pressure behavior of the spinodal line 
shows that data are consistent with the possibility that the two lines 
meet with the same slope. 
If  the pressure $P(T)$ along the spinodal  increases with $T$,  
as usually found in liquids,  the meeting point defines the lowest 
temperature and pressure that can be reached by the liquid in equilibrium. 
These values are $T_I \simeq 73$ K and $P_I \simeq=-360$ MPa.  
\section{Summary and conclusions} 
\label{summary}
In summary,  we have studied the stability domain of the liquid state 
for a simple molecular model for ortho-terphenyl. 
In particular, we have focused on the two limits of stability, 
one provided by the divergence of the structural relaxation times, 
the other one provided by the cavitation of the gas phase.
Both of them have theoretical mean-field bounds, 
the locus at which the configurational entropy vanishes and the locus at 
which the compressibility diverges, respectively. 

To evaluate the configurational entropy, we have developed, along the lines 
of previous work for the same model~\cite{otplungo},  a potential energy 
landscape description of the system free energy, in the framework
of the inherent structure thermodynamic formalism~\cite{stillingerpes}. 
The landscape analysis has required the evaluation of the statistical 
properties of the landscape, which we have quantified in the volume 
dependence of total number, energy distribution, 
and relation between energy depth and shape of the potential energy
landscape basins.

The landscape analysis performed here is limited to volumes such that the 
minimization procedure, used for the evaluation of the inherent structures,
results in an homogeneous structure. We have found that
for volumes larger than the so-called Sastry density~\cite{debenlewis} 
($ V_{Sastry} \simeq 225$ cm$^3$/mol) the system 
always cavitates upon minimization. Cavitation prevents the possibility 
of estimating the statistical properties of the landscape for  
$V \geq V_{Sastry}$.  This poses a serious problem to the application of
landscape approaches in the version where minimizations are performed at 
constant $V$ if the region close to the spinodal curve has to be investigated.  
Indeed,  as shown in Fig.~\ref{fig:locivt}, the region of stability of the
liquid phase extends well beyond $V_{Sastry}$.   
This suggests also that the region close to the spinodal curve is
stabilized by vibrational contributions, which must overcome the destabilizing 
contribution arising from the configurational degrees of freedom 
(as discussed in Sec.~\ref{sec:eos}).  
The stabilization in the vibrational properties appears to be accomplished 
by a softening of the vibrational density of states on approaching $V_{Sastry}$.
We note on passing that, while in the $(T,V)$ ensemble, estimates of the liquid 
free energy in the IS formalism are limited to $V<V_{Sastry}$, formulations 
of the IS formalism in the $(P,T)$ ensemble does not suffer
from such limitation, since the minimization path would not meet any instability 
curve. 

In the framework of the IS formalism, we have estimated the locus at which 
configurational entropy vanishes. The possibility of a finite $T$ at which 
$S_{conf}$ vanishes is encoded in the model selected to represent the data.  
For all densities and temperatures studied in the present work, the Gaussian 
landscape~\cite{heuer00} well represents the data and provides a well defined 
Kauzmann locus, whose location in the phase diagram has been compared with the 
location of the spinodal line.  
We have shown that the spinodal and the $S_{conf}=0$ loci may be extrapolated 
to meet at $V_{Sastry}$ at a finite $T$. Data are also consistent with the 
possibility that, in the $(P,T)$ plane, the two curves are tangent at 
$V_{Sastry}$. These two observations are in agreement with 
the behavior recently predicted for hard and soft spheres complemented 
by a mean field attraction. 
It will be interesting to address in the future the relation between the 
field of stability of the liquid as compared to the field of stability 
of the glass state.
\begin{acknowledgments}
The authors acknowledge support from Miur COFIN 2002 and FIRB 
and INFM-PRA GenFdT and S.~Sastry, P.~G.~Debenedetti and R.~J.~Speedy 
for useful discussions.
\end{acknowledgments}

\begin{thebibliography}{99}
\bibitem{notakauz}
Note that in the present work the Kauzmann locus is not defined as the locus 
of point where the liquid entropy is equal to the crystal entropy,
at variance with the original Ref.~\cite{kauzm}.
\bibitem{notaerrore} 
Note that in previous papers this quantity has been replaced with  
%
\begin{equation} 
f_{harm}(E_{IS},T,V)=k_B T \langle \sum_{i=1}^{6N-3} 
\ln(\beta\hbar\omega_i(E_{IS}))\rangle'.
\label{eq:corr}
\end{equation}
%
Differences between Eqs.~(\ref{eq:fvib}) and (\ref{eq:corr}) 
are of the order of about $1\%$.
Note that data evaluated according to Eq.~(\ref{eq:fvib}) are noisier 
due to errors propagation~\cite{cristianothesis}.
\bibitem{notadistribuzione} 
%
We note that this hypothesis breaks down in the very-low-energy tail, 
where differences between the Gaussian distribution and the actual 
distribution become relevant. As discussed in Ref.~\cite{heuer00}, 
the system Gaussian behavior reflects also some properties of the 
independent subsystems.
\bibitem{notasastry}
Moreover, for $V>V_{Sastry}$,  size effects becomes extremely relevant, 
since the probability of nucleating of a gas bubble during the minimization 
procedure  is a function of the system volume.
\end{thebibliography}
\end{document}